\documentclass[manuscript]{aastex}
\usepackage{emulateapj5}
\usepackage{apjfonts}
\usepackage{epsfig}

\journalinfo{Accepted by The Astrophysical Journal Letters }
\slugcomment{Received 2007 July 1; accepted 2007 December 4}
\shorttitle{Early Growth of Massive Black Holes}
\shortauthors{WANG, CHEN, YAN \& HU} 

\def\soltan{\ifmmode {So\l tan} \else {So\l tan}\ \fi}

\def\calnbh{{\cal N}(\mbh,z)}

\def\pdotm{\langle \dot{m}\rangle}

\def\pmseed{\langle\mbh^{\rm S}\rangle}
\def\pmz{\langle\mbh(z)\rangle}

\def\Psibh{\Psi(M_{\bullet},z)}

\def\rd{{\rm d}}

\def\rhos{\rho_{\bullet}^{\rm S}}

\def\ergs{\ifmmode {\rm erg~ s^{-1}} \else {\rm erg~s^{-1}}\ \fi}
\def\kms{\ifmmode {\rm km~ s^{-1}} \else {\rm km~s^{-1}}\ \fi}

\def\mbh{M_{\bullet}}
\def\mbhc{M_{\bullet}^*}
\def\mgii{\ifmmode Mg {\sc ii} \else Mg {\sc ii}\ \fi}

\def\rhobh{\rho_{\bullet}}
\def\rmd{{\rm d}}

\def\sunm{M_{\odot}}

\def\ergs{${\rm erg~s^{-1}}$}

\def\lax{{$\mathrel{\hbox{\rlap{\hbox{\lower4pt\hbox{$\sim$}}}\hbox{$<$}}}$}}
\def\gax{{$\mathrel{\hbox{\rlap{\hbox{\lower4pt\hbox{$\sim$}}}\hbox{$>$}}}$}}

\begin{document}

\title{Early Growth of Massive Black Holes in Quasars}

\author{Jian-Min Wang\altaffilmark{1}, 
        Yan-Mei Chen\altaffilmark{1,3}, 
	Chang-Shuo Yan\altaffilmark{1,3}, and
	Chen Hu\altaffilmark{1,2,3}}

\altaffiltext{1}{Key Laboratory for Particle Astrophysics, Institute of High Energy Physics, 
                 Chinese Academy of Sciences, 19B Yuquan Road, Beijing 100049, China}

\altaffiltext{2}{National Astronomical Observatories of China, Chinese Academy of Science, 20A Datun
Road, Beijing 100012, China}

\altaffiltext{3}{Graduate School, Chinese Academy of Science, 19A Yuquan Road, Beijing 100049, China}

\begin{abstract}
Episodic activity of quasars is driving growth of supermassive black holes (SMBHs) via accretion of 
baryon gas. In this Letter, we develop a simple method to analyse the duty cycle of quasars up to redshift 
$z\sim 6$ universe from luminosity functions (LFs). We find that the duty cycle below redshift $z\sim 2$ 
follows the cosmic history of star formation rate (SFR) density. Beyond $z\sim 2$, the evolutionary trends 
of the duty cycle are just opposite to that of the cosmic SFR density history, implying the role of feedback 
from black hole activity. With the duty cycle, we get the net lifetime of quasars ($z\le 5$) about 
$\sim 10^9$yrs. Based on the local SMBHs, the mean mass of SMBHs is obtained at any redshifts and their 
seeds are of $10^5\sunm$ at the reionization epoch ($z_{\rm re}$) of the universe through the conservation 
of the black hole number density in comoving frame. We find that primordial black holes ($\sim 10^3\sunm$)
are able to grow up to the seeds via a moderate super-Eddington accretion of $\sim 30$ times of the critical 
rate from $z=24$ to $z_{\rm re}$. Highly super-Eddington accretion onto the primordials is not necessary.
\end{abstract}
\vskip 0.8cm

\section{Introduction}
Accretion of gas onto SMBHs is powering the huge energy output from quasars, 
but SMBH formation and quasar's lifetime remain open in this well-established paradigm. The 
elegant idea, comparing the total accreted mass density during the active phases with the local 
mass density of SMBHs in normal galaxies (\soltan 1982, hereafter \soltan argument), has been 
examined in detail from quasar LF (Chokshi \& Turner 1992; Yu \& Tremaine 2002) and X-ray background 
(Marconi et al. 2004). This generally convinces us that accretion during their episodic activities
is the main source of mass growth. However, the SMBH growth at different redshifts remains open. 

In the present paper, we make an attempt to develop an efficient way to estimate the duty 
cycle for high redshifts, offering a new clue to understand SMBH growth at different redshifts. 
We use the cosmological parameters $H_0=75~{\rm km~s^{-1}~Mpc^{-1}}$,
$\Omega_{\rm M}=0.3$ and $\Omega_{\Lambda}=0.7$ throughout this paper.

\section{Duty Cycle of Quasars}
The SMBH mass function at redshift $z$ in active and dormant galaxies is $\calnbh$,
the duty cycle is defined as a fraction of active black holes to the total, 
\begin{equation}
\delta (\mbh,z)=\frac{\Psibh}{\calnbh},
\end{equation}
where $\mbh$ is black hole mass and $\Psibh$ is the mass function of black holes in quasars at a 
redshift $z$. The averaged duty cycle is $\delta_1(z)=N_{\rm qso}(z)/N_{\rm all}(z)$ in term of 
the number density, where $N_{\rm qso}(z)=\int_{\mbhc}\Psibh \rmd \mbh$, 
$N_{\rm all}(z)=\int_{\mbhc}\calnbh \rmd \mbh$, and $\mbhc$ is the lower mass limit of SMBHs in 
the sample. In the meanwhile, the mass-weighted duty cycle is given by 
$\delta_2(z)=\rhobh(z)/\rhobh^{\rm all}(z)$, where $\rhobh(z)=\int_{\mbhc}\Psibh \mbh \rmd \mbh$ 
and $\rhobh^{\rm all}(z)=\int_{\mbhc}\calnbh\mbh \rmd \mbh$. It has been shown that
$\delta_1(z)=\delta_2(z)\equiv\delta(z)$ in Wang et al. (2006c), namely
\begin{equation}
\delta(z)=\frac{\rhobh(z)}{\rhobh^{\rm all}(z)}.
\end{equation}	
Eq. (2) involves that the mean masses of active and inactive black holes are equal, and then converts 
the number density to mass density ratio (Wang et al. 2006c). It should be noted that 
$\delta(z)$ represents the duty cycle of the major population of SMBHs at a given redshift. 

We get the mass density of black holes from quasar LF if we assume a constant $\dot{m}$ of quasars 
as in literatures (e.g. Marconi et al. 2004, but see evidence for this in Kollmeier et al. 2006).  
The dimensionless accretion rate is defined as $\dot{m}=\dot{M}/\dot{M}_{\rm crit}$, where 
$\dot{M}$ is accretion rates of black holes, $\dot{M}_{\rm crit}=L_{\rm Edd}/c^2$ the critical rate, 
$L_{\rm Edd}=\mbh c^2/t_{\rm Salp}$ the Eddington luminosity,
$t_{\rm Salp}=\sigma_{\rm T}c/4\pi Gm_{\rm p}=0.45$ Gyr the Salpeter time, $\sigma_{\rm T}$  
the Thomson scattering section, $c$ the light speed,
$G$ the gravity constant and $m_{\rm p}$ the proton mass. With the help of
$L_{\rm Bol}=\eta \dot{M}c^2=\eta \dot{m} \mbh c^2/t_{\rm Salp}$, where $\eta$ is the radiative 
efficiency and quasar LF, we have the black hole mass density in active 
galaxies at redshift $z$
\begin{equation}
\rhobh(z)=\frac{1}{\eta\pdotm c^2}\dot{U}(z)t_{\rm Salp},
\end{equation}
where $\pdotm$ is the mean dimensionless accretion rate, and the luminosity density is given by
\begin{equation}
\dot{U}(z)=\int_{L_{\rm Bol}^*} L_{\rm Bol} \Phi(L_{\rm Bol},z)\rd L_{\rm Bol},
\end{equation}
where $\Phi(L_{\rm Bol},z)$ is the bolometric LF, $L_{\rm Bol}$ the 
bolometric luminosity, and $L_{\rm Bol}^*$ its corresponding limit. 
If the mass density of seed black holes of quasars is $\rhos$ at their birth epoch 
($z_{\rm max}$), the mass density of all (active and inactive) black holes is given by
\begin{equation}
\rhobh^{\rm all}(z)=\rhos+\int_z^{z_{\rm max}}\frac{1-\eta}{\eta}\frac{\dot{U}(z)}{c^2}
                    \left(\frac{\rd t}{\rd z}\right)\rd z
                   =\rhos+\frac{1-\eta}{\eta}\frac{U(z)}{c^2},
\end{equation}
where $U(z)=\int_z^{z_{\rm max}}\dot{U}(\rmd t/\rmd z) \rmd z$. The dependence of $\eta$ on 
$z$ can be neglected in eq. (5) as shown by Wang et al. (2006b) from Sloan Digital Sky 
Survey (SDSS) data. We almost know nothing about $\rhos$ at $z_{\rm max}$ 
except for some limited information on the reionization of the universe (Madau et al. 2004).
Inserting (3) and (5) into (2), we have
\begin{equation}
\delta(z)=\frac{\dot{U}(z)t_{\rm Salp}}{\pdotm\left[U_{\rm S}+(1-\eta)U(z)\right]}
         \approx\frac{\dot{U}(z)t_{\rm Salp}}{\pdotm\left(1-\eta\right)U(z)} ,
\end{equation}
where $U_{\rm S}=\eta\rhos c^2$ is the energy density of the seed black holes and
the approximation is valid for $\rhos/\rhobh^{\rm all}(z)\ll 1$.

Information on $\rhos$ can be estimated from the reionization of the universe. The WMAP (Wilkinson
Microwave Anisotropy Probe) detected a large optical depth to Thomson scattering, $\tau_e=0.17\pm 0.04$,
and suggests that the reionization happened at much higher redshift $z_{\rm re}=17\pm 3$ (Spergel et al. 
2003). Madau et al. (2004) suggest that the reionization may be powered by mini-quasars, which have 
a mass density of accreting black holes at least $2\times 10^3\sunm/{\rm Mpc}^3$, which roughly 
agrees with that extrapolated by the luminosity function used below. We take 
$\rhos=2\times 10^3\sunm/{\rm Mpc^3}$ and $\eta=0.1$ in this paper.

We would like to point out followings: 1) the approximation is accurate
enough within a certain redshift $z_c$ when $\rhos$ can be neglected and baryon accretion 
dominates. It breaks when $\delta>1$; 2) the advantage of eq. (6) is that $\delta$ is not sensitive 
to $\eta$; 3) $\delta$ is insensitive, if the specific LF is applied, to the bolometric correction 
factor since it will be canceled on both sides of the numerator and the denominator; 4) $\pdotm$ is 
an observable in principle and seems to be a constant at least between $z=0.3\sim 4$ (Kollmeier et al. 
2006, but see Netzer et al. 2007 for a small high redshift sample). Though the cosmological evolution 
of $\dot{m}$ is poorly understood (Netzer \& Trakhtenbrot 2007), the influence of $\pdotm$ is clear 
in eq. (6). These make eq. (6) robust to calculate $\delta$ for high redshift SMBHs
and accurate enough for low redshift ones.

In this Letter, we use the bolometric LF given by Hopkins et al. (2007). It is combined through
bolometric luminosity correction from a large set of LFs in optical, soft and hard X-rays, and near- 
and middle-IR bands and also covers the fraction of obscured quasars (see also Maiolino et al. 2007; 
M\"uller \& Hasinger 2007 for the fraction of type II quasars). We use the LDDE LF given by eq. (11-16), 
of which parameters are listed in Table 4 in Hopkins et al. (2007) and extrapolate luminosity functions 
beyond $z=6.0$.  We assume $z_{\rm max}=z_{\rm re}$ 
and find that the final results are not sensitive to $z_{\rm max}$. 

Fig 1{\em a} shows $\dot{U}(z)$. There is a break at $z\sim 2$ (also small effects on the duty cycle), 
which is caused by the LF. $\dot{U}$ dramatically drops toward low redshifts and gradually decreases
toward high redshifts. The function $U(z)$ is not plotted here, but its trends are equivalent to that
of the cumulative mass 
density of black holes, whose behaviors can be seen from $\pmz$ according to eq. (9) (shown in 
Fig. 2{\em b}). Quasars with redshifts of $0.3\le z\le 4$
have a mean Eddington ratio of $L_{\rm Bol}/L_{\rm Edd}=0.25$ with a scatter
of 0.3 dex, namely, $\langle\dot{m}\rangle=2.5\eta_{0.1}^{-1}$, where $\eta_{0.1}=\eta/0.1$,
using the empirical reverberation relation (Kollmeier et al. 2006). 
Fig 1{\em b} shows the duty cycle of quasars ranging from $z=0$ to $6$ for 
$\langle\dot{m}\rangle=1, 2.5, 5,10$, respectively. 

From Fig. 1{\em b}, we find that there is a gradual decrease of $\delta$ from high redshifts to 
$z\sim 2$ and then dramatically evolves going down to $10^{-3}\sim 10^{-4}$ in the local universe. 
The duty cycle below $z\le2$ is similar to the results from Wang et al. (2006c). 
Comparing with the cosmic history of SFR density, we find $\delta$ has very similar 
evolutionary trend. Exhaustion of gas leads to lower $\dot{\rho}_{\rm SFR}$, so does the fueling 
gas to SMBHs. This strong evolution is thus regarded as the lack of fueling gas, consequently, most 
SMBHs are then starving. Afterglows of quasars are expected to appear (Wang et al. 2005).

\figurenum{1}
\centerline{\psfig{figure=f1.ps,angle=270,width=8.5cm}} 
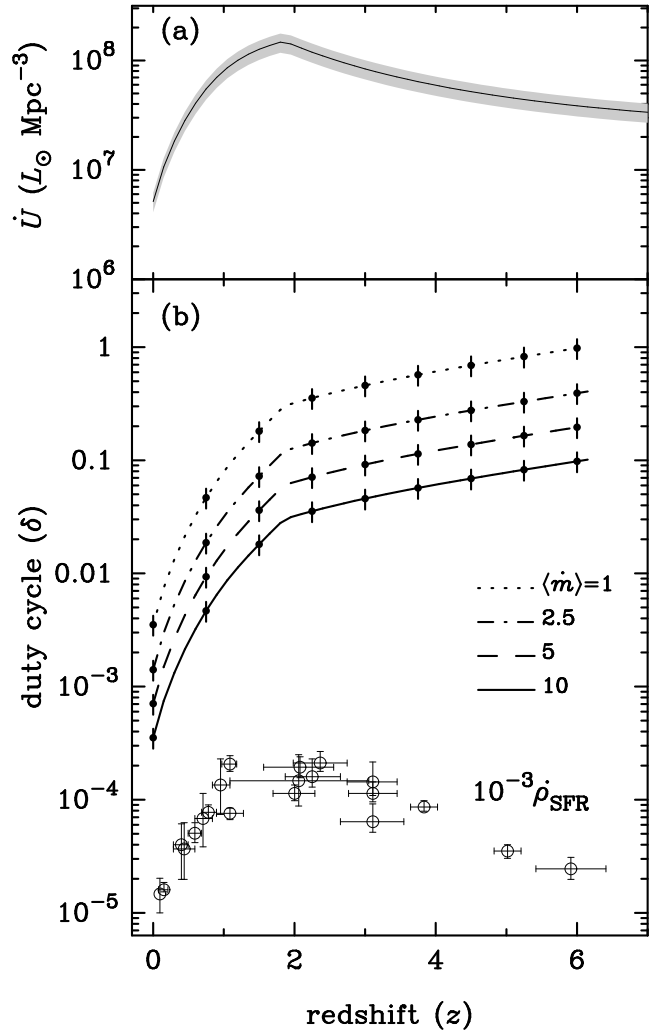
\figcaption{({\em a}) The luminosity density versus redshift based on bolometric luminosity 
function (Hopkins et al. 2007). ({\em b}) The duty cycle versus redshifts. We calculate cases of
$\langle\dot{m}\rangle=10, 5, 2.5, 1$, corresponding to that quasars are radiating at 
$L_{\rm Edd}$, $0.5L_{\rm Edd}$, $0.25L_{\rm Edd}$ and $0.1L_{\rm Edd}$, respectively. The 
error bars are taken from the averaged value of LF 
$\Delta\delta/\delta=\Delta \Phi/\Phi\approx 0.2$. The cosmic history of SFR
density is inserted as open circles taken from Reddy et al. (2007), but scaled by a factor 
of $10^3~\sunm~{\rm yr^{-1}~Mpc^{-3}}$.  }
\label{fig1}
\vglue 0.3cm

Beyond redshift $z\sim 2$, the trends of $\delta$ and $\dot{\rho}_{\rm SFR}$ are just opposite
as shown in Fig 1{\em b}. The present $\delta$ at high redshifts agree very well with the duty 
cycle estimated from clustering of high redshift quasars ($z\ge 2.9$) by Shen et al. (2006). 
The duty cycle gradually decreases from high to low redshifts while $\dot{\rho}_{\rm SFR}$
slowly increases. During this epoch there is enough gas for both star formation and accretion
onto SMBHs. We note that the strong feedback of black hole activities (Schawinski et al. 2006; 
Wang et al. 2007) is driving to blow away fueling gas so as to switch off quasars (Di Matteo et al. 
2005). It has been found by Peng et al. (2006) and 
McLure et al. (2006) that quasars at $z\sim 2$ have stellar mass less than expected from the
local Magorrian relation. This also is indicated by the broken relation between star formation 
and AGN activity at high luminosities (Maiolino et al. 2007), suggesting the SMBHs are 
growing faster than star formation at high redshifts in the presence of strong feedback of 
black hole activity. On the other hand, one can check if the 
star formation still obeys the Kennicutt-Schmidt's law as done in Seyfert galaxies (Wang et al. 
2007). It is expected that further evidence for the evolution of the
feedback is to search from statistics at high redshifts.  Future high spatial resolution 
observations of ALMA (Atacama Large Millimeter Array) 
will directly uncover the detailed nature of feedback.

\section{Accretion and Growth}

\subsection{Lifetime of Quasars}
The net lifetime of quasars (i.e. the total over the Hubble time) is given by
\begin{equation}
t_{\rm QSO}= \int_z^{z_{\rm max}}\delta(z)\left(\frac{\rd t}{\rd z}\right)\rd z,
\end{equation}
where $t$ is the cosmic time. Fig. 2{\em a} shows the net lifetimes for quasars with different
accretion rates. For the rate $\dot{m}=2.5$, we find the typical value is $\le 10^9$yrs
for $z\le 5$. Marconi et al. (2004) obtained a lifetime of a few $10^8$yrs, based on $\dot{m}=10$,
which agrees with our results for the same $\dot{m}$. If a single episodic lifetime is measured
from the transverse proximity effect (Goncalves et al. 2007), the cycles of episodic activities 
can be estimated. The cutoff around $z\sim 15$ in Fig. 2{\em a} 
is caused by setting the birth of quasars at this epoch.

\subsection{Growth of Seed Black Holes from the Primordial }
Neglecting mergers, we have the conserved number density in the co-moving frame as
\begin{equation}
\frac{\rho_{\bullet}^{\rm all}(z)}{\pmz}=\frac{\rhos}{\pmseed}
                                        =\frac{\rho_{\bullet}^0}{\langle\mbh^0\rangle},
\end{equation}
where $\rho_{\bullet}^0$ and $\langle\mbh^0\rangle$ are the SMBH mass density and mean mass in the 
local universe. This is justified by the results from detailed numerical simulations, which show 
the mass contributed from major mergers is only roughly a few percent after $z=10$ epoch (Volonteri 
et al. 2003). The mass density of the local SMBHs
is given by $\rho_{\bullet}^0=4.2\times 10^5\sunm/{\rm Mpc^3}$ (Shankar et al. 2004; Marconi et al. 
2004). We convert the function of dispersion velocity [$\phi(\sigma)$] of early type galaxies (Sheth 
et al. 2003) into the mass function of the SMBHs, $\phi(\mbh)=\phi(\sigma)\rmd \sigma/\rmd \mbh$, 
where $\mbh/\sunm=10^{8.13}(\sigma/200)^{4.02}$ is used (Tremaine et al. 2002). 
The mean mass of local SMBHs is given by 
$\langle \mbh^0\rangle=\int\mbh\phi(\mbh)\rmd \mbh/\int\phi(\mbh)\rmd \mbh=8.9\times 10^7\sunm$.
We thus have the mean mass of black holes at $z$
\begin{equation}
\pmz=\left[\frac{\rho_{\bullet}^{\rm all}(z)}{\rho_{\bullet}^0}\right]\langle\mbh^0\rangle.
\end{equation}
Fig. 2{\em b} shows $\pmz$ as a function of redshifts. If future observations could provide the mean 
mass of SMBH [$\langle\mbh^{\rm obs}(z)\rangle$] for a complete sample at a given redshift, it becomes 
feasible to justify if the non-baryon accretion onto the black holes is necessary by a simple comparison 
of $\langle\mbh^{\rm obs}(z)\rangle$ and $\langle\mbh(z)\rangle$.

From eq. (8), we have the mean mass of seed black holes
\begin{equation}
\pmseed=\left(\frac{\rhos}{\rho_{\bullet}^0}\right)\langle\mbh^0\rangle
        \approx 2.0\times 10^5\sunm,
\end{equation}
where we use $\rhos$ limited by the WMAP. It should be noted that the real $\pmseed$ could be smaller 
than the given by eq. (10) if stars partially ionize the universe. The value of $\pmseed$ agrees well 
with the black hole mass of the mini-quasar model in Madau et al. (2004). How to form such a massive 
seed black hole at $z_{\rm re}$ remains open from the primordial at $z\sim 24$.

\figurenum{2}
\centerline{\psfig{figure=f2.ps,angle=270,width=8.5cm}}
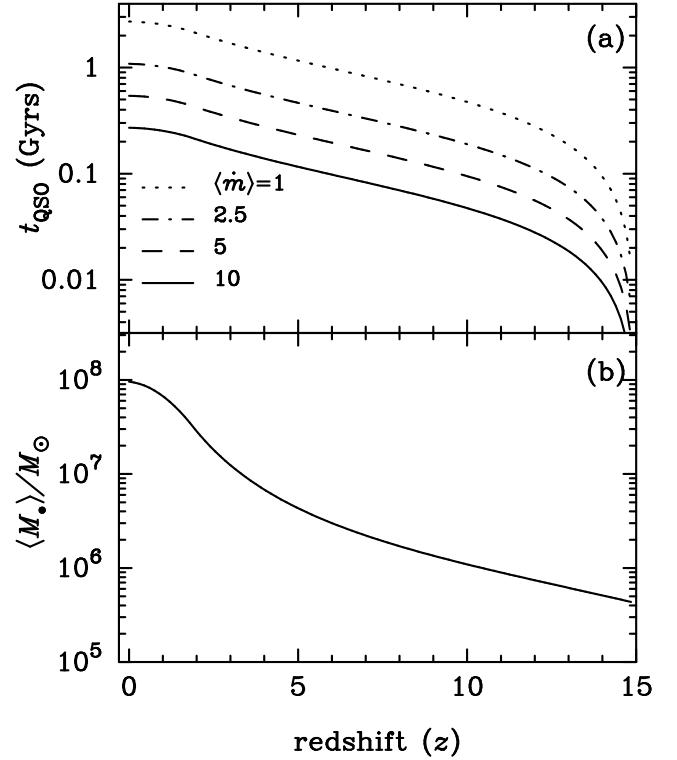
\figcaption{({\em a}) The net lifetime of quasars. ({\em b}) The mean mass of supermassive black holes
at different redshifts.
}
\label{fig2}
\vglue 0.3cm

The primordial black holes can be generally produced by the collapse of population III stars 
(Madau et al. 2001), or collapse of primordial gas clouds (Haehnelt \& Rees 1993; Loeb et al. 1994) 
or gravitational core collapse of relativistic star clusters (Volonteri 2006) with a typical mass 
of $10^2-10^3\sunm$, $10^3-10^6\sunm$ and $10^2-10^4\sunm$, respectively. For a growing primordial 
black hole with a typical mass of $\langle\mbh^{\rm P}\rangle=10^3\sunm$, the necessary accretion 
rate is 
\begin{equation}
\dot{m}_{\rm c}=\left[\frac{t_{\rm Salp}}{(1-\eta)\Delta t}\right]
                \ln\left(\frac{\pmseed}{\langle\mbh^{\rm P}\rangle}\right)\approx 30,
\end{equation}
in the Salpeter growth,
where $\Delta t$ ($\approx 0.09$Gyrs) is the interval between $z=24\sim 17$. This is only moderate
super-Eddington and is realistic in high redshift universe. The photon trapping effects 
make the accretion have low radiative efficiency, but disks of the black holes radiate at a level 
of $L_{\rm Edd}$ (Wang et al. 1999; Wang \& Zhou 1999; Ohsuga et al. 2005).

We note that this conclusion is inconsistent with that in Volonteri \& Rees (2005, hereafter VR05).  
VR05 suggested that a highly super-Eddington accretion onto primordial black holes between 
$z=23\sim 24$ to 
form seed black holes of $10^5\sunm$ in term of the Bondi accretion (see their Fig. 1), and subsequently 
the seeds gradually grow up to be a $10^9\sunm$ SMBH until $z=6$. Using the Bondi rate, we obtain 
$\dot{m}=\dot{m}_0m_{\bullet}^0/\left[1-m_{\bullet}^0\left(t/\tau_0\right)\right]$,
where $m_{\bullet}^0$ is the initial mass of a primordial black hole in units of solar mass,
$\dot{m}_0=G\sunm c\sigma_{\rm T}n_0/c_s^3=0.05~T_{0.8}^{-3/2}n_4$ and 
$\tau_0=c_s^3/4\pi G^2m_{\rm p}n_0\sunm=13.9~T_{0.8}^{3/2}n_4$Gyr, $n_4=n_0/10^4{\rm cm^{3}}$
is the number density of ambient medium, $c_s=\left(kT/m_{\rm p}\right)^{1/2}$ the sound speed of 
the medium, $T_{0.8}=T/8000$K the temperature of the medium and $k$ the Boltzmann constant.
The rate $\dot{m}$ goes to $10^{4\sim 5}$, even infinity for a critical time of $t_c=\tau_0/m_{\bullet}^0$, 
before the Bondi radius is larger than the typical dimension of the primordial clouds (VR05). 
The classical Bondi accretion is valid provided the growth timescale of the accreting 
black hole, $\tau_{\rm BH}=\left(\rmd\ln\mbh/\rmd t\right)^{-1}$ is much longer than the gaseous dynamical
$\tau_{\rm dyn}=R_{\rm Bondi}/c_s$, where $R_{\rm Bondi}$ is the Bondi radius. Under the environment in high
redshift universe, we have 
$\tau_{\rm BH}/\tau_{\rm dyn}=c_s^6/4\pi G^3\mbh^2m_{\rm p}n_0=0.1~T_{0.8}^3M_5^{-2}n_4^{-1}$, 
where $M_5=\mbh/10^5\sunm$. This directly indicates that the Bondi 
approximation is broken for a rapid growth of the primordials.  
A self-consistent treatment of time-dependent Bondi accretion onto a growing black hole
is needed for such a context. 

On the other hand, the rapid growth through super-Eddington accretion with the Bondi rates in VR05
is suffering from strong feedback from outflow or Compton heating (Wang et al. 2006a).
The strength of feedback depends on the angular momentum of the primordial gas accreted onto the 
black hole. A potentially efficient way to avoid the strong feedback is that the primordial gas has 
angular momentum large enough to extend the outer radius of the accretion disk so as to suppress the 
feedback due to the Compton heating. A moderate super-Eddington accretion is thus feasible to form a 
seed black hole from the primordial during the period of $z=24$ to 17.

\section{Conclusions and Discussions}
We develop a convenient way to calculate the duty cycle based on LF, which applies 
to any redshifts. We find that the trends of the duty cycle and the cosmic history of the SFR density
are just opposite when $z>2$. This could be explained by AGN feedback to star formation. With the duty 
cycle, the net lifetime of quasars can be obtained from the quasar luminosity function. The mean mass 
of the seed black holes is up to $2\times 10^5\sunm$ at $z\sim 17$, which is able to grow up from the 
primordial ($\sim 10^3\sunm$) at $z=24$ via moderate super-Eddington accretion. More deeper surveys are 
expected to improve the LFs for more sophisticated investigations of growth of black holes in high 
redshift universe.

\acknowledgements{We are very grateful to the referee for profoundly thoughtful reports that greatly 
improve the paper. J.M.W. thanks J. P. Ostriker and D. N. C. Lin for helpful discussions, and H. Netzer
for many motivated conversations. We appreciate the stimulating discussions among the members of IHEP 
AGN group. The research is supported by NSFC and CAS via NSFC-10325313, 10733010 and 10521001, and 
KJCX2-YW-T03, respectively.}


\begin{thebibliography}{}
\bibitem[]{402}Chokshi, A. \& Turner, E. L. 1992, MNRAS, 259, 421
\bibitem[]{403}Di Matteo, T., Springel, V. \& Hernquist, L. 2005, \nat, 443, 604
\bibitem[]{}Goncalves, T. S., Steidel, C. C. \& Pettini, M. 2007, \apj, astro-ph/0711.4113
\bibitem[]{405}Haehnelt, M. G. \& Rees, M. J. 1993, \mnras, 263, 168
\bibitem[]{407}Hopkins, H., Richards, G. T. \& Herquist, L., 2007, ApJ, 654, 731 
\bibitem[]{410}Kollmeier, J. A. et al. 2006, \apj, 648, 128
\bibitem[]{411}Loeb, A. \& Rasio, F. A. 1994, \apj, 432, 52
\bibitem[]{414}McLure, R. J. et al. 
2006, MNRAS, 368, 1395
\bibitem[]{416}Madau, P. \& Rees, M. J. 2001, \apj, 551, L27
\bibitem[]{417}Madau, P. et al. 
2004, ApJ, 604, 484
\bibitem[]{419}Maiolino, A. et al. 
2007, A\&A, 468, 979
\bibitem[]{421}Marconi, A. et al. 
2004, \mnras, 351, 169
\bibitem[]{423}M\"uller, A. \& Hasinger, G. 2007, astro-ph/0708.0942
\bibitem[]{}Netzer, H., Lira, P. \& Trakhtenbrot, B. 2007, \apj, astro-ph/0708.3787
\bibitem[]{424}Netzer, H. \&  Trakhtenbrot, B. 2007, \apj, 654, 754
\bibitem[]{426}Ohsuga, K., Mori, M., Nakamoto, T. \& Mineshige, S. 2005, ApJ, 628, 368 
\bibitem[]{427}Peng, C. et al. 2006, ApJ, 640, 114
\bibitem[]{}Reddy, N. A. et al. 2007, \apjs, in press (astro-ph/0706.4091)
\bibitem[]{432}Schawinski, K. et al. 2006, Nature, 442, 888
\bibitem[]{433}Shankar, F., Salucci, P., Granato, G. L., De Zotti, G. \& Danese, L. 2004, MNRAS, 354, 1020
\bibitem[]{434}Shen, Y., Strauss, M. A., et al., 2006, AJ, 133, 2222
\bibitem[]{435}Sheth, R. K. et al. 2003, ApJ, 594, 225
\bibitem[]{437}So\l tan, A. 1982, \mnras, 200, 115
\bibitem[]{438}Spergel, D. N. et al. 2003, ApJS, 148, 175
\bibitem[]{439}Tremaine, S. et al. 2002, ApJ, 574, 740
\bibitem[]{440}Volonteri, M., Haardt, F. \& Madau, P. 2003, \apj, 582, 559
\bibitem[]{441}Volonteri, M. \& Rees, M. J. 2005, ApJ, 633, 624
\bibitem[]{442}Volonteri, M. 2006, astro-ph/0602630  
\bibitem[]{443}Wang, J.-M., Chen, Y.-M. \& Hu, C. 2006a, \apj, 637, L85
\bibitem[]{444}Wang, J.-M., Chen, Y.-M., Ho, L. C. \& McLure, R. J. 2006b, ApJ, 642, L111
\bibitem[]{445}Wang, J.-M., Chen, Y.-M. \& Zhang, F. 2006c, \apj, 647, L17
\bibitem[]{446}Wang, J.-M., Chen, Y.-M., Yan, C.-S., Hu, C. \& Bian, W.-H. 2007, ApJ, 661, L143
\bibitem[]{447}Wang, J.-M., Szuszkiewicz, E., Lu, F.-J. \& Zhou, Y.-Y., 1999, ApJ, 522, 893
\bibitem[]{448}Wang, J.-M., Yuan, Y.-F. \& Ho, L. C. 2005, ApJ, 625, L5
\bibitem[]{449}Wang, J.-M. \& Zhou, Y.-Y. 1999, ApJ, 516, 420
\bibitem[]{451}Yu, Q. \& Tremaine, S. 2002, MNRAS, 335, 965
\end{thebibliography}
\end{document}